\PassOptionsToPackage{table,xcdraw}{xcolor}
\documentclass{article}

\usepackage[preprint]{neurips_2024}

\usepackage[utf8]{inputenc} 
\usepackage[T1]{fontenc}    
\usepackage{hyperref}       
\usepackage{url}            
\usepackage{booktabs}       
\usepackage{amsfonts}       
\usepackage{nicefrac}       
\usepackage{microtype}      
\usepackage{colortbl}        
\usepackage{newfloat}
\usepackage{tcolorbox}
\usepackage{listings}
\tcbuselibrary{listings, breakable}
\usepackage{pifont}
\newcommand{\cmark}{\ding{51}} 
\newcommand{\xmark}{\ding{55}} 
\usepackage{booktabs}
\usepackage{enumitem}
\usepackage{multirow}
\usepackage{array}
\usepackage{tablefootnote}
\usepackage{arydshln}
\usepackage{threeparttable}
\usepackage{subcaption}
\usepackage{natbib}
\title{WenetSpeech-Yue: A Large-scale Cantonese Speech Corpus with Multi-dimensional Annotation}

\author{%
  Longhao Li$^{1}$\thanks{Equal contribution.}\\
  \And
  Zhao Guo$^{1}$\footnotemark[1] \\
  \And
  Hongjie Chen$^{2}$ \\
  \And
  Yuhang Dai$^{1}$ \\
  \And
  Ziyu Zhang$^{1}$ \\
  \And
  Hongfei Xue$^{1}$ \\
  \And
  Tianlun Zuo$^{1}$ \\
  \And
  Chengyou Wang$^{1}$ \\
  \And
  Shuiyuan Wang$^{1}$ \\
  \And
  Jie Li$^{2}$ \\
  \And
  Jian Kang$^{2}$ \\
  \And
  Xin Xu$^{3}$ \\
  \And
  Hui Bu$^{3}$ \\
  \And
  Binbin Zhang$^{4}$ \\
  \And
  Ruibin Yuan$^{5}$ \\
  \And
  Ziya Zhou$^{5}$ \\
  \And
  Wei Xue$^{5}$ \hspace{1.5em} Lei Xie$^{1}$\thanks{Corresponding author.} \\[0.5em]
  $^{1}$Audio, Speech and Language Processing Group (ASLP@NPU),\\
Northwestern Polytechnical University \\
  $^{2}$Institute of Artificial Intelligence (TeleAI), China Telecom \\
  $^{3}$Beijing AISHELL Technology Co., Ltd. \\
  $^{4}$WeNet Open Source Community \\
  $^{5}$Hong Kong University of Science and Technology \\[0.5em]
  \texttt{lhli@mail.nwpu.edu.cn, gzhao@mail.nwpu.edu.cn, lxie@nwpu.edu.cn} \\[0.5em]
  \textbf{Project page:} \url{https://github.com/ASLP-lab/WenetSpeech-Yue}
}

\begin{document}

\maketitle

\setcounter{footnote}{0}
\begin{abstract}
The development of speech understanding and generation has been significantly accelerated by the availability of large-scale, high-quality speech datasets. Among these, ASR and TTS are regarded as the most established and fundamental tasks. However, for Cantonese (Yue Chinese), spoken by approximately 84.9 million native speakers worldwide, limited annotated resources have hindered progress and resulted in suboptimal ASR and TTS performance. To address this challenge, we propose WenetSpeech-Pipe, an integrated pipeline for building large-scale speech corpus with multi-dimensional annotation tailored for speech understanding and generation. It comprises six modules: Audio Collection, Speaker Attributes Annotation, Speech Quality Annotation, Automatic Speech Recognition, Text Postprocessing and Recognizer Output Voting, enabling rich and high-quality annotations. Based on this pipeline, we release WenetSpeech-Yue, the first large-scale Cantonese speech corpus with multi-dimensional annotation for ASR and TTS, covering 21,800 hours across 10 domains with annotations including ASR transcription, text confidence, speaker identity, age, gender, speech quality scores, among other annotations. We also release WSYue-eval, a comprehensive Cantonese benchmark with two components: WSYue-ASR-eval, a manually annotated set for evaluating ASR on short and long utterances, code-switching, and diverse acoustic conditions, and WSYue-TTS-eval, with base and coverage subsets for standard and generalization testing. Experimental results show that models trained on WenetSpeech-Yue achieve competitive results against state-of-the-art (SOTA) Cantonese ASR and TTS systems, including commercial and LLM-based models, highlighting the value of our dataset and pipeline. 
\end{abstract}

\section{Introduction}

Recent advances in speech understanding and generation have been largely fueled by the availability of large-scale, diverse, and richly annotated datasets. Core tasks such as ASR and TTS exemplify how model performance across diverse linguistic and acoustic conditions hinges on this very foundation. For instance, advanced ASR and TTS systems, such as FireRedASR~\cite{fireredasr}, Whisper~\cite{whisper}, and VALL-E~\cite{valle}, have achieved significant performance breakthroughs by leveraging the advantages of large, diverse corpora. To meet the growing demand for high-quality resources, scalable data pipelines like GigaSpeech2~\cite{gigaspeech2}, WenetSpeech~\cite{wenetspeech}, and Emilia-Pipe~\cite{emilla} series have streamlined the construction of large, multilingual, and multi-domain corpora, thereby facilitating the efficient development of ASR and TTS systems.

Despite these advances, Cantonese, as a Chinese dialect with high practical usage and linguistic complexity, remains severely under-resourced. On one hand, Cantonese is spoken by over 84.9 million people across mainland China, Hong Kong, Macau, and global Chinese communities~\cite{yue_scarity}. Its rich tone system of nine tones in six categories, coexistence of literary and colloquial forms, and frequent code-switching with English pose unique modeling challenges. On the other hand, its cultural distinctiveness demands speech systems with high robustness, emotional expressiveness, and stylistic diversity, which current resources fail to adequately support.

\begin{table*}[t]
\centering
\resizebox{\textwidth}{!}{%
\begin{tabular}{lccccccc}
\toprule
\textbf{Dataset} & \textbf{Task} & \textbf{Duration (hours)} & \textbf{Long Audio} & \textbf{Code-switch} & \textbf{Multi-domain} & \textbf{Multi-label} \\
\midrule
Guangzhou Daily Use & ASR & 4.06 & \xmark & \xmark & \xmark & \xmark \\
Guangzhou Cabin & ASR & 5.00 & \xmark & \xmark & \xmark & \xmark \\
Mixed Cantonese and English & ASR & 34.8 & \xmark & \cmark & \xmark & \xmark \\
Common-Voice yue & ASR & 203 & \xmark & \xmark & \xmark & \xmark \\
Common-Voice zh-HK & ASR & 108 & \xmark & \xmark & \xmark & \xmark \\
MDCC & ASR & 73.6 & \xmark & \xmark & \cmark & \xmark \\
ZoengJyutGaai-Storytelling & TTS & 188.25 & \xmark & \xmark & \xmark & \xmark \\
\midrule
WenetSpeech-Yue & ASR/TTS/Others & 21,800 & \cmark & \cmark & \cmark & \cmark \\
\bottomrule
\end{tabular}%
}
\caption{Comparison with existing Cantonese speech datasets.}
\end{table*}

Publicly available Cantonese corpora are often limited in scale, style, and label diversity. Projects like Common Voice~\cite{commonvoice} and MDCC~\cite{MDCC} rely heavily on manual annotation and offer only small datasets. Evaluation sets are typically composed of short utterances and lack coverage of complex linguistic phenomena. Moreover, most corpora provide only speech-text alignment, with little to no speaker attributes or acoustic quality metadata, severely limiting their use in self-supervised learning, style modeling, and multi-task training. As a result, mainstream ASR and TTS systems perform poorly on Cantonese tasks and exhibit weak generalization to real-world scenarios.

To address these gaps, we introduce WenetSpeech-Pipe, a modular and automated pipeline designed for building large-scale Cantonese datasets. It integrates multiple quality-enhancement strategies and metadata enrichment techniques to ensure the resulting corpus is both diverse and richly annotated, supporting a wide range of modeling objectives. Built upon this pipeline, we construct WenetSpeech-Yue, the largest and most comprehensive open-source Cantonese speech corpus to date, spanning over 21,800 hours across eleven domains. 
To facilitate rigorous evaluation of both ASR and TTS systems, we further release WSYue-eval, a dedicated evaluation suite comprising two subsets: WSYue-ASR-eval for ASR and WSYue-TTS-eval for TTS. Each subset is curated to cover a wide range of linguistic and acoustic scenarios, supporting comprehensive and reliable assessment under diverse conditions.
Experimental results show that ASR and TTS models trained on WenetSpeech-Yue achieve performance comparable to or exceeding SOTA systems on multiple test sets.
These contributions significantly advance Cantonese speech understanding and generation, addressing critical resource gaps and enabling more robust and expressive speech systems.



Our contributions can be summarized as follows:
\begin{enumerate}
    \item We propose WenetSpeech-Pipe, a large-scale, multi-domain, multi-label data pipeline tailored for both speech understanding and generation in Cantonese.
    \item We release WenetSpeech-Yue, a 21,800-hour large-scale Cantonese speech corpus with rich multi-dimensional annotations—currently the largest open-source resource for Cantonese speech research.
    \item We release WSYue-eval, a comprehensive Cantonese benchmark comprising WSYue-ASR-eval, a human-annotated test set for ASR, and WSYue-TTS-eval for TTS.
    \item We demonstrate that ASR and TTS models trained on WenetSpeech-Yue achieve SOTA performance across multiple benchmarks.
\end{enumerate}

\section{Related Work}

\subsection{Large-scale Speech Corpora and the Scarcity of Cantonese Resources}
Recent years have seen growing efforts in constructing large-scale open-source speech datasets across languages and modalities. GigaSpeech2~\cite{gigaspeech2} adopts a self-iterative label refinement strategy to develop multilingual corpora for Southeast Asian languages (Thai, Indonesian, and Vietnamese), automating collection, transcription, and alignment. Multilingual LibriSpeech~\cite{librispeech}(MLS), derived from LibriVox audiobooks, offers approximately 451,000 hours of read speech, with 445,000 hours in English and 6,000 in seven other languages. WenetSpeech~\cite{wenetspeech} curates 22,435 hours of Chinese speech audio from YouTube and podcasts, combining with refined transcripts via a well-designed pipeline. WenetSpeech4TTS~\cite{wenetspeech4tts} further refines this data for speech synthesis by adjusting segment length, filtering for speaker consistency, and enhancing audio quality, resulting in 12,800 hours of high-quality TTS-ready speech. Emilia introduces a large-scale multilingual speech generation dataset of over 101,000 hours across six languages, supporting both TTS training and synthesis research.


Despite these advances, publicly available Cantonese corpora remain limited in both scale and linguistic diversity. For example, Mozilla’s Common Voice~\cite{commonvoice} provides around 311 hours of validated Cantonese read speech.
The MDCC~\cite{MDCC} corpus contains 73.6 hours of high-quality audiobook recordings from Hong Kong, spanning domains such as philosophy, education, and lifestyle. The ZoengJyutGaai-Storytelling Dataset\footnote{https://huggingface.co/datasets/CanCLID/zoengjyutgaai} offers 112.54 hours of expressive single-speaker speech. These datasets are often constrained to read speech, narrow domain coverage, and lack of rich metadata—posing challenges for building robust models for real-world Cantonese speech applications.

\subsection{Cantonese ASR and TTS Models}

Despite recent progress, Cantonese-specific speech modeling remains under-resourced compared to major languages. On the ASR front, SenseVoice~\cite{funaudiollm}, trained on 300,000 hours of multilingual speech including 9,600 hours of Cantonese, is regarded as the SOTA Cantonese ASR system. Whisper-large-v3~\cite{whisper}, a general-purpose multilingual model trained on over 5 million hours of speech, also demonstrates strong cross-lingual generalization, achieving competitive performance on Cantonese without dialect-specific tuning. More recently, TeleASR~\cite{telespeechpt}, a dialect-aware model pretrained on 300,000 hours of unlabeled audio and fine-tuned across 30 Chinese dialects, has demonstrated robust recognition capabilities in regional languages including Cantonese.


On the synthesis front, open-source TTS models have increasingly extended support to Cantonese and other dialects. CosyVoice2~\cite{cosyvoice2}, which enables zero-shot voice cloning and multilingual synthesis, includes Cantonese among supported languages such as Mandarin, English, Japanese, and Korean. Meanwhile, Step-Audio-TTS-3B~\cite{step-audio}, a lightweight model within the Step-Audio framework, leverages knowledge distillation from a larger model to support controllable speech generation, including emotion, speaking rate, dialectal variation (e.g., Cantonese, Sichuanese).

\section{WenetSpeech-Pipe}

\begin{figure*}[t]
\centering
\includegraphics[width=1\textwidth]
{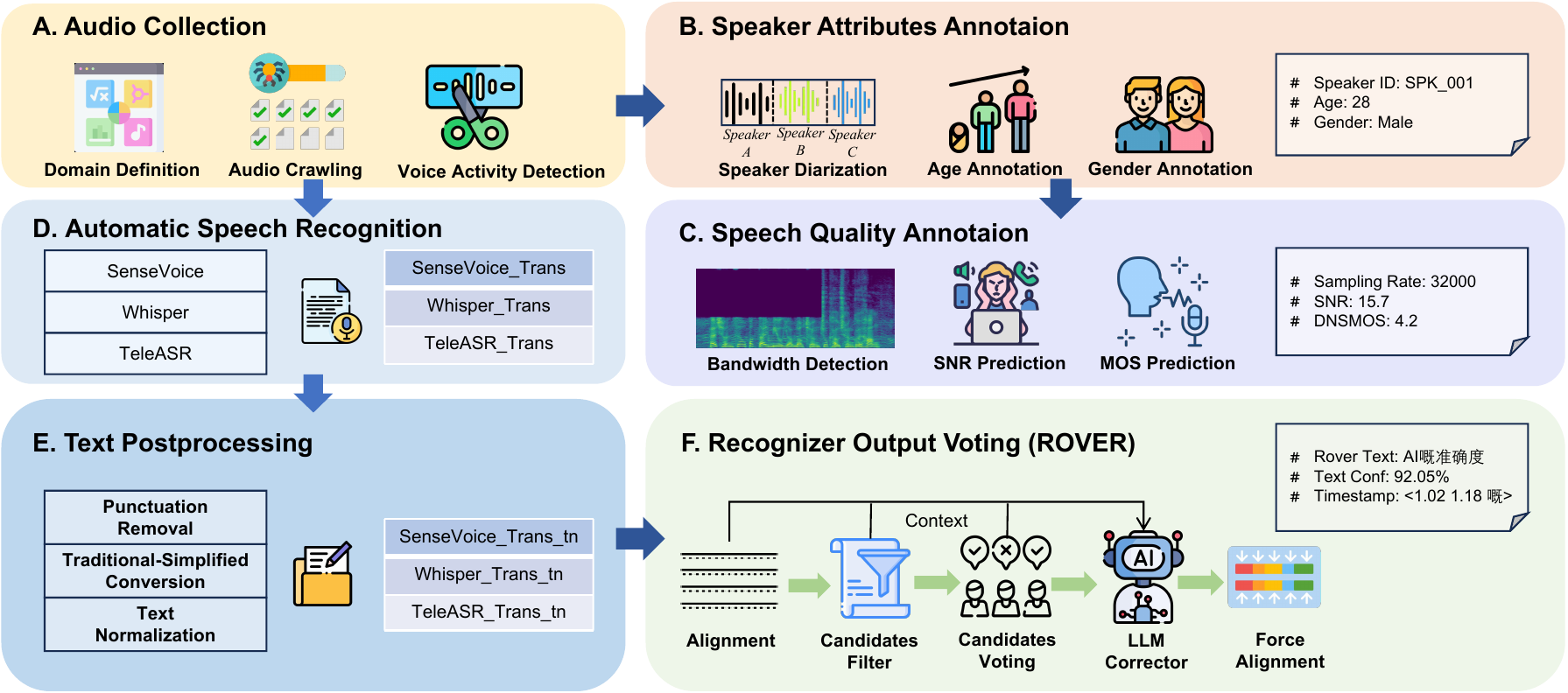} 
\caption{An overview of the WenetSpeech-Pipe processing pipeline. 
}
\label{WS_pipe}

\end{figure*}


Our proposed WenetSpeech-Pipe framework, as illustrated in Figure~\ref{WS_pipe}, comprises six modules: (A) Audio Collection, (B) Speaker Attributes Annotation, (C) Speech Quality Annotation, (D) Automatic Speech Recognition, (E) Text Postprocessing, and (F) Recognizer Output Voting.

\noindent\textbf{Audio Collection.} 
WenetSpeech-Pipe begins with the large-scale collection of in-the-wild speech data spanning diverse domains, such as storytelling, drama, commentary, vlogs, food, entertainment, news, and education.
Raw recordings are typically long-form, often tens of minutes to over an hour, which makes them unsuitable for direct model training or alignment. To generate utterance-level data suitable for downstream transcription and quality assessment, the audio is automatically segmented into short clips using a voice activity detection (VAD) module, producing a large and domain-diverse pool of segmented audio clips ready for subsequent processing.

\noindent\textbf{Speaker Attribute Annotation.} 
To enrich the dataset with speaker-level metadata for multi-speaker modeling and style-aware synthesis, WenetSpeech-Pipe incorporates a Speaker Attributes Annotation stage. 
First, speaker diarization is performed using the pyannote toolkit~\cite{pyannote}, which assigns local speaker labels to short audio segments from the same source, providing intra-recording speaker separation. 
Second, both age and gender are estimated for each segment using the Vox-Profile~\cite{voxprofile}, providing speaker attribute annotations. 
This process produces utterance-level segments annotated with speaker identity, age, and gender, forming a multi-dimensional metadata that facilitates both supervised and style-controllable speech modeling.

\noindent\textbf{Speech Quality Annotation.} 
To ensure that the curated audio supports high-fidelity speech generation tasks such as TTS and voice conversion (VC), WenetSpeech-Pipe incorporates a comprehensive speech quality assessment stage. Each audio segment is assessed through three complementary approaches. First, noise levels are characterized using Brouhaha~\cite{Brouhaha}, which produces segment-level signal-to-noise ratio (SNR) annotations. 
Second, perceptual quality is measured with DNSMOS~\cite{dnsmos}, a non-intrusive model that predicts the mean opinion score (MOS), which reflects human-perceived speech clarity and naturalness.
Finally, spectral characteristics are analyzed via bandwidth detection, which estimates the effective upper frequency and spectral coverage of each recording, complementing nominal sampling-rate metadata. Together, these multi-dimensional assessments generate a structured quality annotation for each segment, providing both quantitative scores (SNR and DNSMOS) and a reliable spectral reference (sampling rate) for downstream high-fidelity speech processing.

\noindent\textbf{Automatic Speech Recognition.} 
Single ASR systems often exhibit systematic biases and error patterns due to architectural constraints~\cite{etai}, training data limitations, or domain mismatch. To mitigate these issues and improve transcription reliability, we adopt a multi-system ensemble approach that leverages diverse recognition paradigms. Specifically, each audio segment is independently transcribed using three high-performance Cantonese ASR systems: the open-source models SenseVoice and Whisper, and the commercial solution TeleASR. These systems differ in architecture, training data, and optimization objectives—enabling complementary error profiles and diverse linguistic hypotheses. The output of this module consists of three parallel transcriptions per utterance, forming the foundation for subsequent fusion and refinement. These multi-hypothesis outputs serve as the primary input to the Recognizer Output Voting stage, where consensus-based alignment and voting yield a more accurate and robust final transcription.

\noindent\textbf{Text Postprocessing.} 
To ensure reliable cross-system alignment and effective integration of multi-source transcriptions, it is essential to standardize the output formats from different ASR systems. The raw transcriptions from the aforementioned ASR systems exhibit significant variations in character sets (traditional vs. simplified Chinese), inclusion of non-lexical tags (e.g., \texttt{[laughter]}), and formatting inconsistencies for numerals and code-switched text. These discrepancies can impede accurate fusion and consensus formation during subsequent processing stages.
Therefore, we apply a text post-processing pipeline to all transcription streams. This pipeline converts traditional Chinese characters to simplified form using the \texttt{OpenCC}\footnote{https://github.com/BYVoid/OpenCC} tool, removes all punctuation marks and special symbols, standardizes numerical expressions and date formats through rule-based rewriting, and inserts whitespace between Cantonese and English words to facilitate bilingual modeling. By applying these steps sequentially, we generate cleaned and canonical transcriptions that are consistent across the three systems. These standardized outputs serve as robust input representations for the ROVER module, ensuring that differences in surface form do not interfere with phonetic or lexical alignment during fusion.
\noindent\textbf{Recognizer Output Voting.} While Text Postprocessing unifies the surface forms of transcriptions across multiple ASR systems, persistent variations remain in lexical selection, word segmentation, and phonetic representation. To generate a unified, high-accuracy reference transcription, we adopt the Recognizer Output Voting Error Reduction (ROVER) framework~\cite{rover}, a fusion strategy based on multi-system voting to enhance transcription accuracy.


In our implementation, we extend the standard ROVER pipeline to handle the linguistic characteristics of Cantonese better. First, transcriptions after text normalization from the aforementioned ASR systems, are aligned using dynamic programming. To ensure robustness against outlier hypotheses, we introduce a candidate filtering module that computes the edit distance between each system's output and the average transcription of the other two. Outputs exceeding a predefined threshold are excluded from voting. At each aligned position, the most frequently occurring word is selected, and the average voting frequency across all positions is recorded as the utterance-level text confidence score. We extend the voting paradigm to Cantonese pinyin by implementing a pronunciation-specific confidence measure, operating in parallel with the character-level voting to reinforce phoneme consistency.

To further improve transcription accuracy, we employ an LLM, Qwen3-4B~\cite{qwen3}, to perform minimal, context-aware refinements on the consensus output. The LLM considers all original ASR hypotheses as contextual references and applies only necessary corrections to grammar, word choice, or named entities, preserving the integrity of the spoken content.

Finally, we perform character-level forced alignment between the refined transcription and the original audio using a pre-trained acoustic model. This yields precise timestamps for each character, enabling fine-grained speech processing and supporting downstream tasks.

\section{WenetSpeech-Yue}

\subsection{Dataset}





\noindent\textbf{Metadata.} Metadata is stored in a single JSON file. The metadata fields include \textit{audio path}, \textit{duration}, 
\textit{text confidence}
, \textit{speaker identity}, \textit{SNR}, \textit{DNSMOS}, \textit{age}, \textit{gender}, and \textit{character-level timestamps}. These fields are extensible, and additional metadata tags may be incorporated in the future.

\begin{figure}[t]
\centering
\includegraphics[width=0.65\columnwidth]{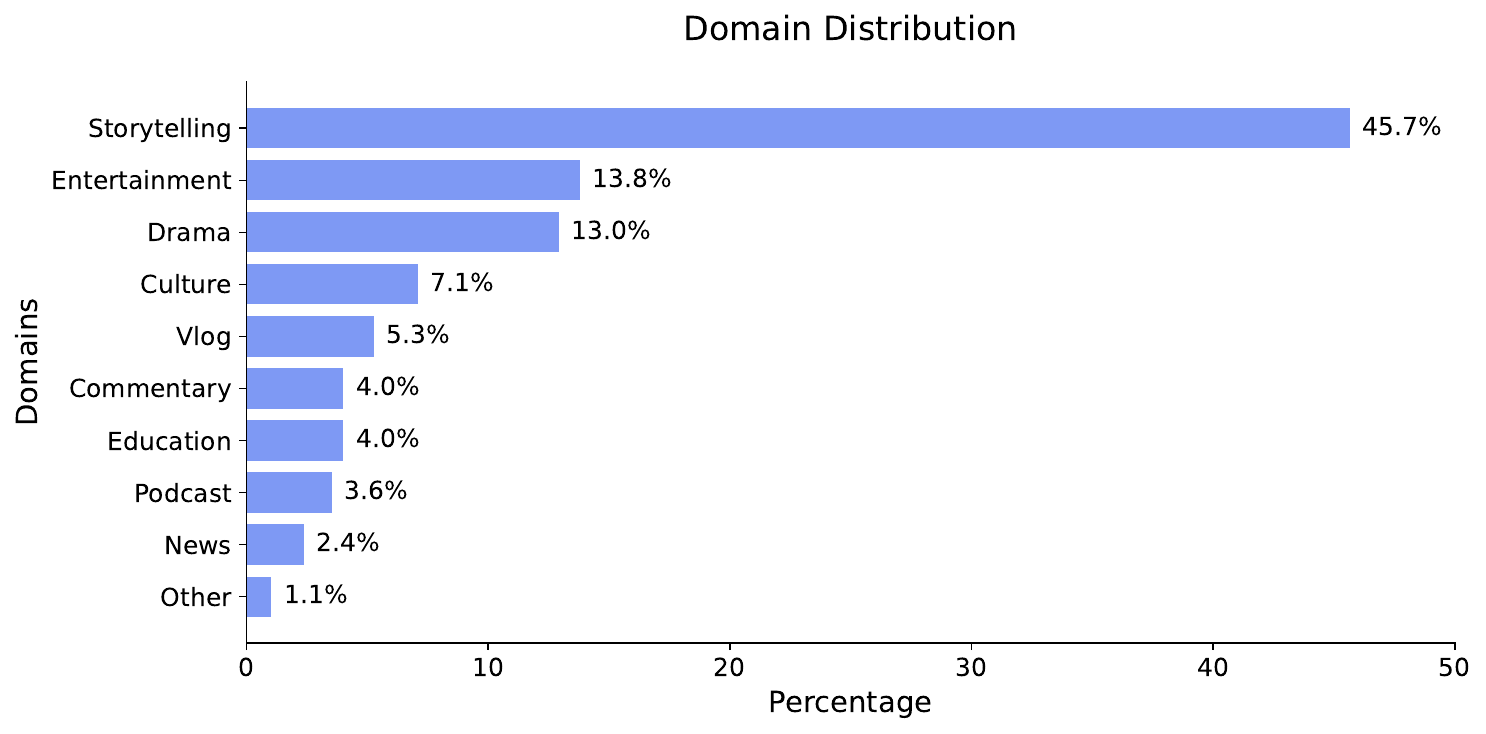} 
\caption{Domain distribution of WenetSpeech-Yue.}
\label{fig_domain}
\end{figure}

\noindent\textbf{Domains.}
The domain of the WenetSpeech-Yue corpus is classified into ten categories: \textit{Storytelling}, \textit{Entertainment}, \textit{Drama}, \textit{Culture}, \textit{Vlog}, \textit{Commentary}, \textit{Education}, \textit{Podcast}, \textit{News}, and \textit{Others}. The distribution of these domains is illustrated in Figure~\ref{fig_domain}.

\begin{figure*}[h]
\centering

\begin{subfigure}[b]{0.31\textwidth}
    \includegraphics[width=\linewidth]{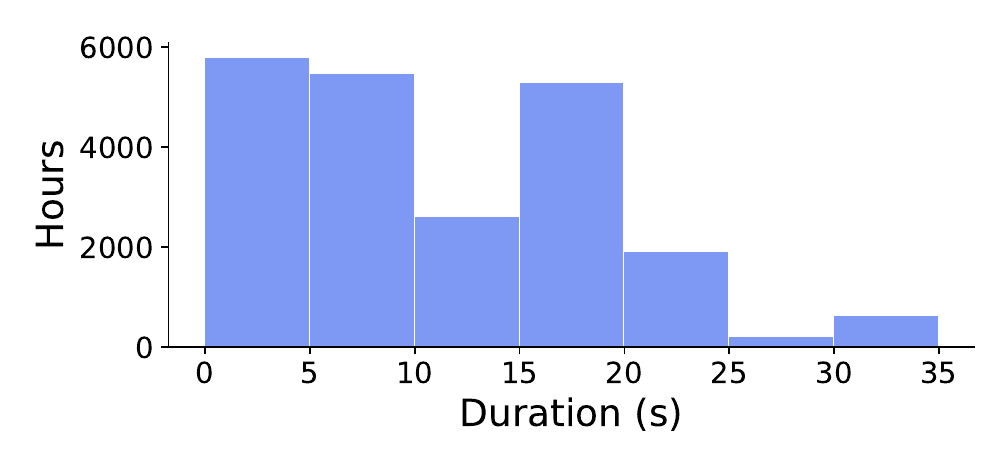}
    \caption{Duration Distribution}
    \label{fig:duration}
\end{subfigure}
\begin{subfigure}[b]{0.31\textwidth}
    \includegraphics[width=\linewidth]{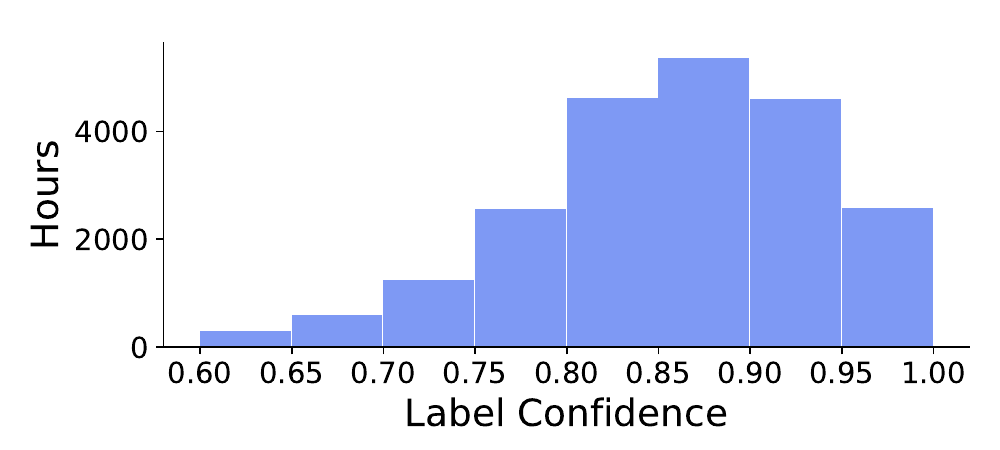}
    \caption{Label Confidence Distribution}
    \label{fig:confidence}
\end{subfigure}
\begin{subfigure}[b]{0.31\textwidth}
    \includegraphics[width=\linewidth]{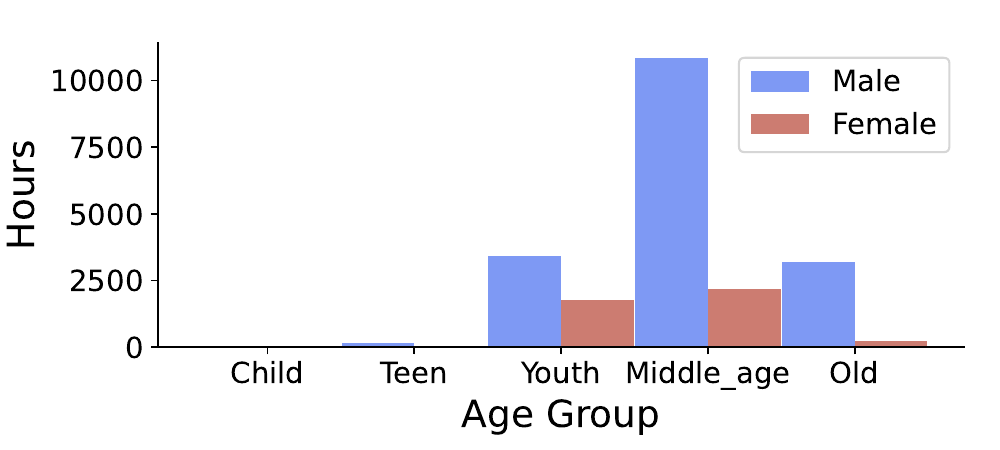}
    \caption{Age-Gender Distribution}
    \label{fig:age_gender}
\end{subfigure}


\begin{subfigure}[b]{0.31\textwidth}
    \includegraphics[width=\linewidth]{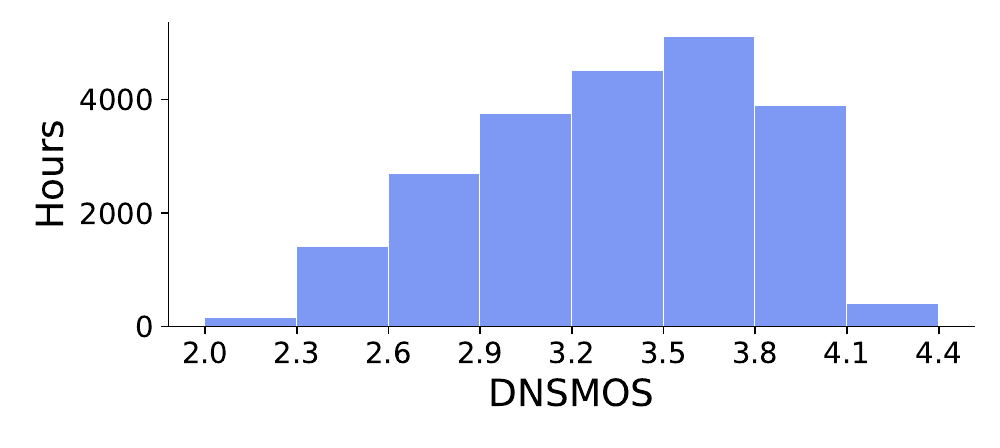}
    \caption{DNSMOS Distribution}
    \label{fig:dnsmos}
\end{subfigure}
\begin{subfigure}[b]{0.31\textwidth}
    \includegraphics[width=\linewidth]{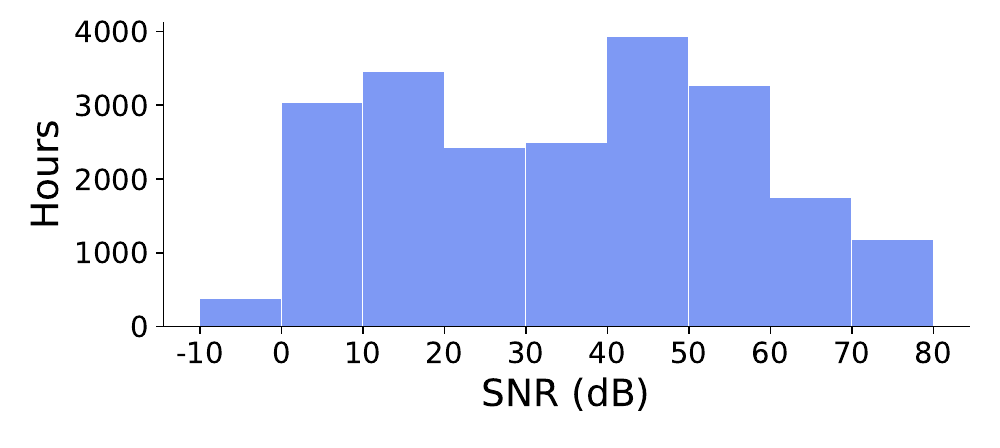}
    \caption{SNR Distribution}
    \label{fig:snr}
\end{subfigure}
\begin{subfigure}[b]{0.31\textwidth}
    \includegraphics[width=\linewidth]{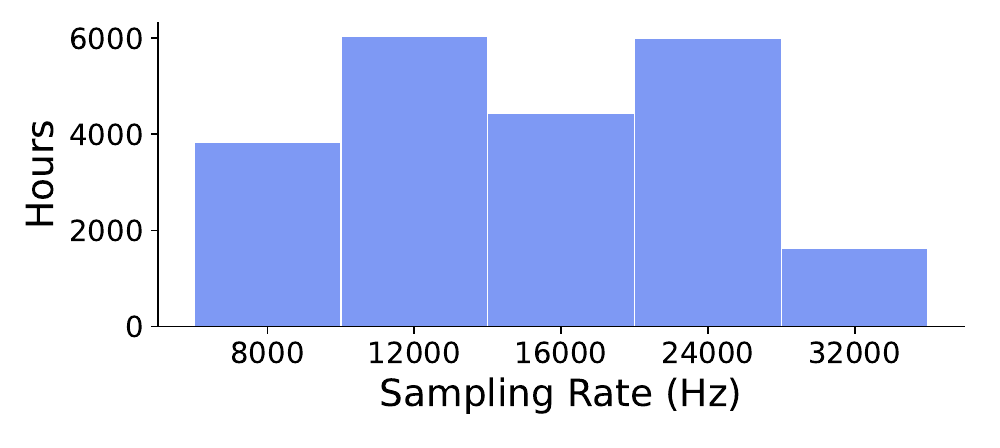}
    \caption{Sampling Rate Distribution}
    \label{fig:band}
\end{subfigure}
\caption{Visualization of statistical analysis for the WenetSpeech-Yue corpus. }
\label{fig:six_subfigs}
\end{figure*}

\noindent\textbf{Duration.}
WenetSpeech-Yue contains a total of 21,800 hours of audio, including both short and long recordings, with an average duration of 11.40 seconds per audio segment. The major distribution is shown in Figure~\ref{fig:duration}.

\noindent\textbf{Confidence.} In this work, we retain only labels with a text confidence score above 0.6. Based on the confidence score, we partition the data into three subsets: \textit{strong labels} (confidence $> 0.9$, 6,771.43 hours), \textit{moderate labels} ($0.8 < $ confidence $\leq 0.9$, 10,615.02 hours), and \textit{weak labels} ($0.6 < $ confidence $\leq 0.8$, 4,488.13 hours).
Detailed distribution is shown in Figure~\ref{fig:confidence}.




\noindent\textbf{Speech Quality.} As illustrated in Figure~\ref{fig:six_subfigs}, we evaluated the audio quality of the WenetSpeech-Yue dataset. Specifically, Figure~\ref{fig:dnsmos} presents \textit{DNSMOS} scores spanning 2.0 to 4.4; Figure~\ref{fig:snr} depicts \textit{SNR} values varying from $-5$ to $80$~dB; and Figure~\ref{fig:band} shows the distribution of \textit{sampling rate}, which range from 8,000 to 32,000~Hz. To ensure suitability for generative tasks, we filtered the dataset by retaining samples with \textit{DNSMOS} greater than 2.5 and \textit{SNR} above 25~dB, resulting in a subset of 12,000 hours high-quality utterances for TTS applications. This filtering strategy can be further adapted for downstream tasks such as vocoder, codec, and voice conversion.



\noindent\textbf{Speaker Attributes.} As shown in Figure~\ref{fig:age_gender}, we present the distribution of age and gender in the WenetSpeech-Yue corpus. 
The corpus is predominantly composed of male speakers, especially in the \textit{Middle\_age} (50.6\%) group, while female speakers are relatively underrepresented in all age groups.

\subsection{Benchmark}
\label{sec:benchmark}

To address the unique linguistic characteristics of Cantonese, we propose \textbf{WSYue-eval}, a comprehensive benchmark encompassing both ASR and TTS tasks. This integrated evaluation framework is specifically tailored to assess model performance across critical dimensions of Cantonese language processing.

\noindent\textbf{ASR Benchmark.} As a representative task of speech understanding, we developed the \textbf{WSYue-ASR-eval} test set for the ASR task. The WSYue-ASR-eval set is annotated through multiple rounds of manual labeling and includes tags such as text transcription, emotion, age, and gender. Based on differences in audio duration, WSYue-ASR-eval is divided into two subsets, \textit{Short} and \textit{Long}, as shown in Table~\ref{tab:asr_evaluation_sets}, to enable comprehensive evaluation of Cantonese speech of varying lengths. In addition, WSYue-ASR-eval covers a wide range of Cantonese usage scenarios, including Cantonese-English code-switching and multi-domain conditions.

\begin{table}[htbp]
\caption{WSYue-ASR-eval Subsets}
\centering
\begin{tabular}{lccc}
\toprule
\textbf{Set} & \textbf{Duration} & \textbf{Speakers} & \textbf{Hours} \\
\midrule
Short & 0--10 s & 2861 & 9.46 \\
Long & 10--30 s & 838 & 1.97 \\
\bottomrule
\end{tabular}
\label{tab:asr_evaluation_sets}
\end{table}

\noindent\textbf{TTS Benchmark.} We introduce \textbf{WSYue-TTS-eval}, a benchmark specifically designed for zero-shot Cantonese TTS evaluation. WSYue-TTS-eval comprises two subsets: \textit{Base} and \textit{Coverage}. The \textit{Base} subset contains 1,000 prompt-text pairs sampled from the CommonVoice dataset, enabling assessment of model performance on real-world data distributions. However, as CommonVoice primarily consists of daily conversational data, its coverage of diverse domains and linguistic phenomena is limited. To address this, the \textit{Coverage} subset combines manually curated and LLM-generated texts. This subset spans a wide range of domains, including daily life, news, entertainment, and poetry, and incorporates various Cantonese linguistic phenomena such as polyphonic characters, tone sandhi, code-switching, proper nouns, numerals, and other challenging cases. This comprehensive design enables rigorous evaluation of model generalization and robustness across diverse and complex scenarios. 

\section{Experiments}

\subsection{ASR Task}

\noindent\textbf{Experimental Setup.}
To assess the effectiveness of the WenetSpeech-Yue corpus and quantify its contribution to model performance, we conducted ASR experiments with two model categories, traditional architectures without large language models and LLM-augmented hybrids, denoted as \texttt{w/o LLM} and \texttt{w/ LLM}, respectively.
The \texttt{w/o LLM} group includes \texttt{U2pp-Conformer-Yue}, a U2pp-Conformer~\cite{u2pp,delayed_kd} trained from scratch to maximize the dataset’s supervision signal, \texttt{Whisper-medium-Yue}, a Whisper-medium model fine-tuned with a low learning rate for efficient Cantonese adaptation, and \texttt{SenseVoice-small-Yue}, a fine-tuned SenseVoice-small~\cite{funaudiollm} variant serving as a strong small-scale Cantonese baseline.
The \texttt{w/ LLM} is showcased by \texttt{U2pp-Conformer-LLM-Yue}, a Conformer–LLM hybrid with the U2pp-Conformer encoder connected to the Qwen3-4B via a lightweight adapter module. 
All models adopt a two-stage training strategy: an initial stage using mixed medium- and high-confidence labels for rapid convergence, followed by a fine-tuning stage on high-confidence labels to maximize transcription accuracy. This setup both reduces training cost and directly reflects the dataset’s quality impact. 


%

\noindent\textbf{Baselines.}
To demonstrate the effectiveness of our training on the WenetSpeech-Yue dataset, we benchmarked against several prominent and competitive Cantonese ASR systems as shown in Table~\ref{tab:asr_results}. 

\noindent\textbf{ASR Test Sets.}
To comprehensively assess the generalization capabilities of various ASR models, we employ a diverse array of Cantonese test sets categorized into three groups: (1) \textbf{In-house collections}, including \textit{Dialogue} for conversational speech analysis, and \textit{Reading} for read speech evaluation; (2) \textbf{Open-source resources}, consisting of the \textit{Common Voice} series (yue and zh-HK)~\cite{commonvoice}, the multi-domain \textit{MDCC}\footnote{https://github.com/HLTCHKUST/cantonese-asr} dataset, \textit{Daily\_Use}\footnote{https://huggingface.co/datasets/AlienKevin/guangzhou-daily-use-speech} (containing 4.06~hours of transcripted daily conversations), and \textit{Commands}\footnote{https://huggingface.co/datasets/AlienKevin/guangzhou-cabin-speech} (featuring 5~hours of in-vehicle command recordings); and (3) Our proposed \textbf{\textit{WSYue-ASR-eval} benchmark} including \textit{Short} and \textit{Long} for short and long-sentence evaluation respectively.

\begin{table}[t]
\caption{ASR Results (MER\%) on various Cantonese test sets for our models and comparison models. The gray bars represent the baseline models, while the green bars represent our proposed models. The dashed lines are used to separate models of different sizes. The best results are highlighted in bold, while the second-best results are underlined.}
\centering
\scriptsize
\setlength{\tabcolsep}{2pt}
\renewcommand{\arraystretch}{1.0}
\resizebox{\textwidth}{!}{%
\begin{tabular}{lc*{9}{c}}
\toprule
\multirow{2}{*}{Model} & {\#Params} & \multicolumn{2}{c}{In-House} & \multicolumn{5}{c}{Open-Source} & \multicolumn{2}{c}{WSYue-ASR-eval} \\
\cmidrule(lr){3-4} \cmidrule(lr){5-9} \cmidrule(lr){10-11}
 & (M) & Dialogue & Reading & yue & zh-HK & MDCC & Daily\_Use & Commands & Short & Long \\
\midrule
\textbf{w/o LLM} & & & & & & & & & & \\
\rowcolor{gray!10} Paraformer~\cite{paraformer} & 220 & 83.22 & 51.97 & 70.16 & 68.49 & 47.67 & 79.31 & 69.32 & 73.64 & 89.00 \\
\rowcolor{gray!10} SenseVoice-small~\cite{funaudiollm} & 234 & 21.08 & \underline{6.52} & 8.05 & \textbf{7.34} & 6.34 & 5.74 & \underline{6.65} & 6.69 & 9.95 \\
\rowcolor{green!10} SenseVoice-small-Yue & 234 & 19.19 & 6.71 & 6.87 & 8.68 & \underline{5.43} & 5.24 & 6.93 & 5.23 & 8.63 \\
\rowcolor{gray!10} Dolphin-small~\cite{dolphin} & 372 & 59.2& 7.38 & 39.69 & 51.29 & 26.39 & 7.21 & 9.68 & 32.32 & 58.20 \\
\rowcolor{green!10} U2pp-Conformer-Yue & 130 & \textbf{16.57} & 7.82 & 7.72 & 11.42 & 5.73 & 5.73 & 8.97 & \underline{5.05} & 8.89 \\
\hdashline

\rowcolor{gray!10}TeleASR~\cite{telespeechpt} & 700 & 37.18 & 7.27 & 7.02 & \underline{7.88} & 6.25 & 8.02 & \textbf{5.98} & 6.23 & 11.33 \\
\rowcolor{gray!10} Whisper-medium~\cite{whisper} & 769 & 75.50 & 68.69 & 59.44 & 62.50 & 62.31 & 64.41 & 80.41 & 80.82 & 50.96 \\
\rowcolor{green!10} Whisper-medium-Yue & 769 & 18.69 & 6.86 & \underline{6.86} & 11.03 & 5.49 & \underline{4.70} & 8.51 & \underline{5.05} & \underline{8.05} \\
\hdashline
\rowcolor{gray!10} FireRedASR-AED-L~\cite{fireredasr} & 1100 & 73.70 & 18.72 & 43.93 & 43.33 & 34.53 & 48.05 & 49.99 & 55.37 & 50.26 \\
\rowcolor{gray!10} Whisper-large-v3~\cite{whisper} & 1550 & 45.09 & 15.46 & 12.85 & 16.36 & 14.63 & 17.84 & 20.70 & 12.95 & 26.86 \\
\hline
\textbf{w/ LLM} & & & & & & & & & & \\
\rowcolor{gray!10} Qwen2.5-Omni-3B\cite{qwen2.5_omni} & 3000 & 72.01 & 7.49 & 12.59 & 11.75 & 38.91 & 10.59 & 25.78 & 67.95 & 88.46 \\
\rowcolor{gray!10} Kimi-Audio~\cite{kimi_audio} & 7000 & 68.65 & 24.34 & 40.90 & 38.72 & 30.72 & 44.29 & 45.54 & 50.86 & 33.49 \\
\rowcolor{green!10} FireRedASR-LLM-L~\cite{fireredasr} & 8300 & 73.70 & 18.72 & 43.93 & 43.33 & 34.53 & 48.05 & 49.99 & 49.87 & 45.92 \\
\rowcolor{green!10} U2pp-Conformer-LLM-Yue & 4200 & \underline{17.22} & \textbf{6.21} & \textbf{6.23} & 9.52 & \textbf{4.35} & \textbf{4.57} & 6.98 & \textbf{4.73} & \textbf{7.91} \\
\bottomrule
\end{tabular}
}
\label{tab:asr_results}
\end{table}

%


\noindent\textbf{ASR Results and Analysis.} 
We use the Mixed Error Rate (MER), which calculates errors at the character level for Chinese and the word level for English, as the evaluation metric to compare models trained on WenetSpeech-Yue and baselines.
As shown in Table~\ref{tab:asr_results}, the experimental results reveal several consistent observations: (1) Across all model scales—including the small, medium, and w/LLM configurations—our models achieve the best performance on most evaluation sets; (2) Within the small-scale group, both \texttt{SenseVoice-small-Yue} and \texttt{U2pp-Conformer-Yue} achieve competitive results, with \texttt{SenseVoice-small-Yue} outperforming all baselines despite having the smallest size, indicating that our corpus substantially enhances efficiency in low-capacity settings; (3) Within the \texttt{w/o LLM} category, both \texttt{U2pp-Conformer-Yue} and \texttt{Whisper-medium-Yue} surpass the large scale baselines; (4) Within the \texttt{w/ LLM} group, \texttt{U2pp-Conformer-LLM-Yue}, consistently attains the-state-of-the-art (SOTA) accuracy. 
Collectively, these observations highlight that our propose WenetSpeech-Yue not only improves overall performance but also maximizes model potential across different parameter regimes, validating its utility for both traditional and LLM-enhanced ASR paradigms.

\begin{table}[h]
\caption{Model performance on WSYue-ASR-eval subsets (MER\%)}
\centering
\setlength{\tabcolsep}{3pt}
\renewcommand{\arraystretch}{1.1}
\resizebox{0.51\textwidth}{!}{
\begin{tabular}{llcc}
\toprule
\multirow{2}{*}{Model} & \multirow{2}{*}{Subset} & \multicolumn{2}{c}{WSYue-ASR-eval} \\
\cmidrule(lr){3-4}
 & & Short & Long \\
\midrule
\multirow{2}{*}{Whisper-medium-Yue} 
 & Stage 1 & 7.27 &  11.19 \\
 & Stage 2 & 5.05 &  8.05 \\
\hdashline

\multirow{2}{*}{U2pp-Conformer-Yue}
 & Stage 1 & 7.62 & 12.01 \\
 & Stage 2 & 5.05 &  8.89 \\
\hdashline

\multirow{2}{*}{U2pp-Conformer-LLM-Yue}
 & Stage 1 & 6.81 & 10.75 \\
 & Stage 2 & 4.73 & 7.91 \\
\bottomrule
\end{tabular}
}
\label{tab:full_results}
\end{table}
Table~\ref{tab:full_results} reports the impact of our two-stage training strategy. Stage 1, trained on the mixed-confidence dataset, already achieves very competitive Cantonese ASR performance, while Stage 2 fine-tuning on high-confidence data yields significant gains across both test sets of WSYue-ASR-eval. These observations confirm that high-confidence labels are the primary driver of performance improvements. 
We believe that retaining confidence information is essential because it facilitates flexible training strategies, allowing high-confidence subsets to drive fine-tuning, while carefully leveraging low-confidence segments can improve model robustness in semi-supervised or domain-adaptive scenarios.


\subsection{TTS Task}

\noindent\textbf{Experimental Setup.}
To evaluate the effectiveness of WenetSpeech-Yue for speech synthesis, we adopt a transfer learning approach on two pretrained TTS models: Llasa-1B and CosyVoice2. Llasa-1B is a zero-shot TTS model pretrained on 250,000 hours of Mandarin and English speech. Both models are further fine-tuned on the TTS subset of WenetSpeech-Yue to adapt them to the linguistic and acoustic characteristics of Cantonese.


\noindent\textbf{Baselines.}
We compare our proposed methods, \texttt{Llasa-1B-Yue} and \texttt{CosyVoice2-Yue}, with several zero-shot baselines: (1) \texttt{Llasa-1B} (without fine-tuning on our dataset); (2) \texttt{CosyVoice2} (without fine-tuning on our dataset); and (3) \texttt{Step-Audio-TTS-3B}. In addition, we include \texttt{Edge-TTS}, a commercial system with single fixed speaker, as a reference for high-quality synthesis.

\noindent\textbf{Evaluation Metrics.}



We employ both objective and subjective metrics to comprehensively assess model performance. For objective evaluation, we use the WSYue-TTS-eval benchmark, which consists of two test subsets: \textit{Base} and \textit{Coverage}. We report MER to measure intelligibility and speaker similarity (SIM) to evaluate voice consistency on both subsets. 
For MER, we transcribe the generated audio using \texttt{U2pp-Conformer-Yue}, our state-of-the-art Cantonese ASR model, and compute MER against the reference text. SIM is measured using Wespeaker~\cite{wespeaker}, which calculates the similarity between speaker embeddings of the synthetic and reference audio.
Additionally, we adopt UTMOSv2~\cite{utmosv2} to assess the overall quality of the synthesized audio.



For subjective evaluation, we randomly select Thirty samples from the \textit{base} test set and Twenty samples from the \textit{coverage} test set. Ten native Cantonese speakers are recruited to evaluate the five TTS systems. We conduct Mean Opinion Score (MOS) evaluations using a 5-point scale and report results with 95\% confidence intervals. The MOS evaluation covers three aspects: intelligibility (I-MOS), speaker similarity (S-MOS), and accent nativeness (A-MOS). Specifically, I-MOS measures the intelligibility and consistency of the synthesized speech with respect to the input text, S-MOS evaluates the similarity between the synthesized and reference speakers, and A-MOS assesses the nativeness of the pronunciation.





\noindent\textbf{TTS Results and Analysis.}

\begin{table}[t]
\caption{TTS performance comparison on WSYue-TTS-eval using both objective and subjective metrics.}
\centering
\small
\setlength{\tabcolsep}{2pt}
\begin{threeparttable}
\resizebox{1\textwidth}{!}{
\begin{tabular}{lcccccccc}
\toprule
\multirow{2}{*}{Model} & \multicolumn{2}{c}{Base} & \multicolumn{2}{c}{Coverage} & \multirow{2}{*}{UTMOSv2 $\uparrow$} & \multirow{2}{*}{I-MOS $\uparrow$} & \multirow{2}{*}{S-MOS $\uparrow$} & \multirow{2}{*}{A-MOS $\uparrow$} \\
\cmidrule(lr){2-3} \cmidrule(lr){4-5} & MER (\%) $\downarrow$ & SIM $\uparrow$ & MER (\%) $\downarrow$ & SIM $\uparrow$ \\
\midrule
\rowcolor{gray!10} Llasa-1B           & 53.31          & 0.732 & 43.68 & 0.754 & 2.360 & 2.60 $\pm$ 1.01 & 3.05 $\pm$ 0.87 & 2.32 $\pm$ 0.98 \\
\rowcolor{gray!10} Step-Audio-TTS-3B  & 27.79          & 0.762 & 24.25 & 0.781 & 2.496 & 3.22 $\pm$ 0.70 & 3.14 $\pm$ 0.58 & 2.82 $\pm$ 0.69\\
\rowcolor{gray!10} CosyVoice2         & 14.38          & 0.812 & 13.74 & 0.826 & 2.989 & 3.72 $\pm$ 0.50 & 3.52 $\pm$ 0.36 & 3.22 $\pm$ 0.60\\
\rowcolor{gray!10} Edge-TTS$\dagger$  & \textbf{8.30}           & -        & \textbf{9.27} & -        & 2.997 & 4.12 $\pm$ 0.28 & -               & 3.48 $\pm$ 0.56\\
\midrule
\rowcolor{green!10} Llasa-1B-Yue       & 10.89          & 0.762 & 12.78 & 0.772 & 2.696 & 4.30 $\pm$ 0.23 & \textbf{4.11} $\pm$ \textbf{0.37} & \textbf{4.34} $\pm$ \textbf{0.34}\\
\rowcolor{green!10} Cosyvoice2-Yue     & \textbf{10.33}          & \textbf{0.821} & \textbf{9.49}  & \textbf{0.834} & \textbf{3.021} & \textbf{4.45} $\pm$ \textbf{0.16} & 3.78 $\pm$ 0.53 & 4.21 $\pm$ 0.27\\
\bottomrule
\end{tabular}
}
\begin{tablenotes}
\footnotesize
\item[$\dagger$] Commercial system with single fixed speaker, and speaker similarity is not considered.
\end{tablenotes}
\end{threeparttable}
\label{tab:tts-results}
\end{table}

For objective metrics, both \texttt{Llasa-1B-Yue} and \texttt{CosyVoice2-Yue} achieve substantial improvements over their pretrained counterparts. In particular, \texttt{CosyVoice2-Yue} attains the lowest MER among all systems, reducing the error rate to $10.33\%$ on the \textit{base} set and $9.49\%$ on the \textit{coverage} set, while also achieving the highest SIM scores (0.821 and 0.834), indicating superior speaker similarity. \texttt{Llasa-1B-Yue} also significantly reduces MER (10.89\% and 12.78\%) compared to \texttt{Llasa-1B}, and maintains competitive SIM values. In addition, UTMOSv2 scores show that both fine-tuned models generate more natural-sounding speech than zero-shot baselines.


In terms of subjective evaluation, \texttt{CosyVoice2-Yue} achieves the highest intelligibility (I-MOS: \textbf{4.45} $\pm$ \textbf{0.16}), while \texttt{Llasa-1B-Yue} outperforms all other systems in speaker similarity (S-MOS: \textbf{4.11} $\pm$ \textbf{0.37}) and accent nativeness (A-MOS: \textbf{4.34} $\pm$ \textbf{0.34}). Notably, although \texttt{CosyVoice2-Yue} obtains higher objective SIM scores, its perceived speaker similarity is lower than that of \texttt{Llasa-1B-Yue}. This may be attributed to the in-context learning inference approach in \texttt{Llasa-1B-Yue}, which leads to more natural prosody and speaking style, thus improving perceived speaker similarity. Compared to zero-shot baselines, both fine-tuned models significantly improve all MOS metrics, and even though \texttt{Edge-TTS} achieves a lower MER, its I-MOS is lower than that of \texttt{CosyVoice2-Yue} or \texttt{Llasa-1B-Yue}, likely because it produces more mechanical and less natural-sounding speech. Overall, these results confirm the effectiveness of WenetSpeech-Yue for improving Cantonese speech synthesis.

\subsection{Conclusion}
In this work, we present WenetSpeech-Yue, the largest and most comprehensive open-source Cantonese speech corpus to date, together with WenetSpeech-Pipe, a scalable and modular data processing pipeline designed for building high-quality speech corpus with multi-dimensional annotations.
We also release the WSYue-eval benchmark to support rigorous ASR and TTS evaluation. Built upon WenetSpeech-Yue, we obtain several ASR/TTS models with SOTA performance consistently. 
We believe that WenetSpeech-Yue and WenetSpeech-Pipe will serve as valuable resources for the community, facilitating future research on multi-domain Cantonese speech understanding and generation.

\setlength{\bibhang}{0pt}

\bibliography{reference}


\newpage
\appendix

\section*{Appendix}
\section{Meta Data Example}

\begin{figure}[h]
\centering
\includegraphics[width=0.6\columnwidth]{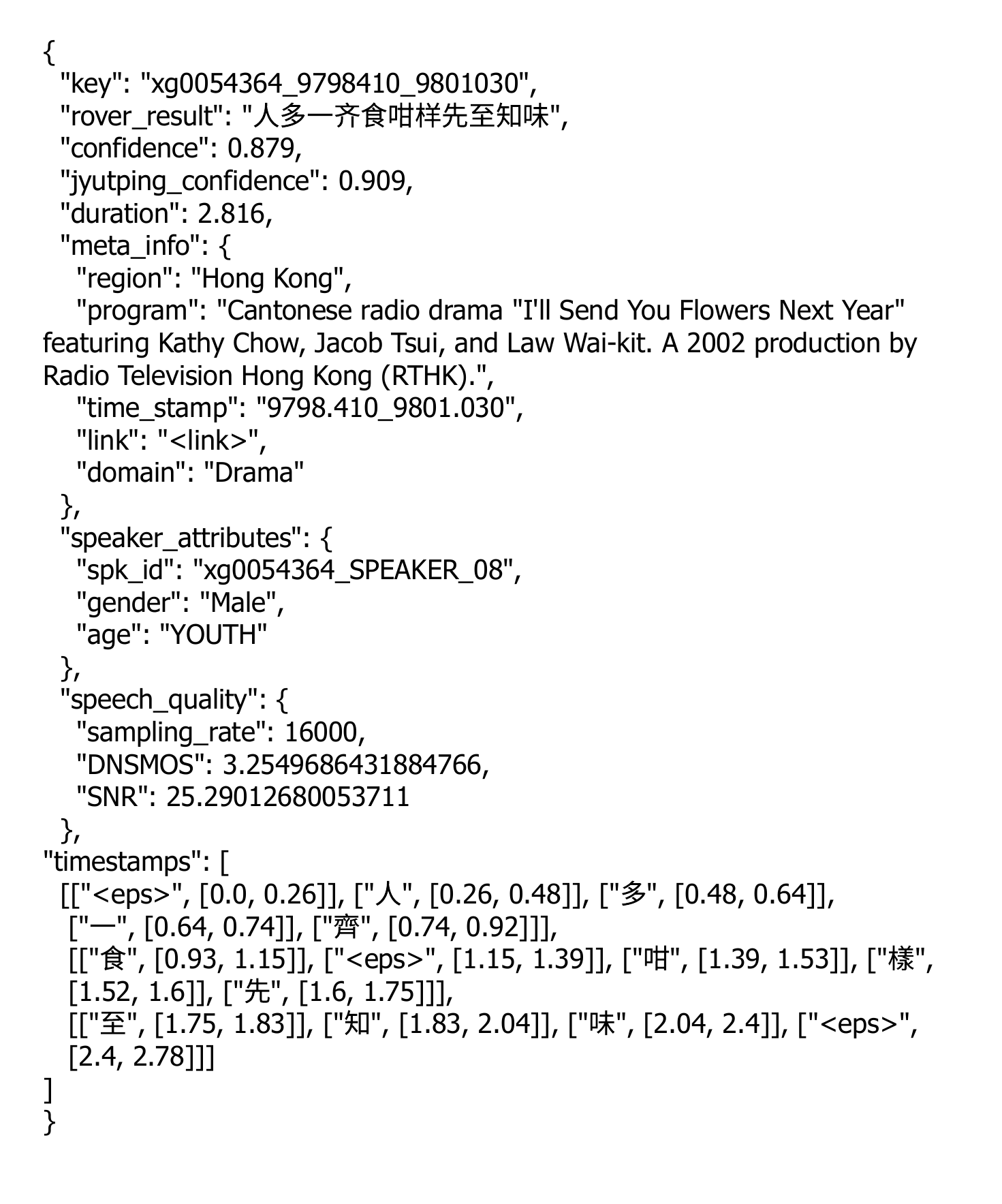} 
\caption{Sample annotations in JSON format.}
\label{meta_data}
\end{figure}

We store all audio metadata in a standardized JSON format. Core fields include \texttt{utt\_id} (unique identifier for each audio segment), \texttt{rover\_result} (rover result of three ASR transcriptions), \texttt{confidence} (confidence score of text transcription), \texttt{jyutping\_confidence} (confidence score of Cantonese pinyin transcriptions), and \texttt{duration} (audio duration). Speaker attributes contain parameters such as \texttt{speaker\_id}, \texttt{gender}, and \texttt{age}. Audio quality assessment metrics include professional measures such as \texttt{sample\_rate}, \texttt{DNSMOS}, and \texttt{SNR}. Timestamp information (\texttt{timestamp}) precisely records start and end times (\textit{start/end}). Extended metadata under the \texttt{meta\_info} field encompasses \texttt{program} (program name), \texttt{region} (geographical information), \texttt{link} (original content link), and \texttt{domain} (domain classification). The metadata example is shown in Figure~\ref{meta_data}.

\section{WenetSpeech-Pipe Details}\label{appendix:pipe detail}

\subsection{Text Postprocessing}

To ensure consistent transcription formats across different ASR systems, we propose an integrated text post-processing framework that performs four key operations: punctuation removal through regular expression matching to eliminate symbolic characters, Traditional-to-Simplified Chinese conversion using the \texttt{OpenCC}\footnote{https://github.com/yichen0831/opencc-python} library, text normalization handled by the \texttt{an2cn}\footnote{https://github.com/Ailln/cn2an} numerical conversion tool, and proper word spacing implemented with the \texttt{Pangu}\footnote{https://github.com/vinta/pangu.py} tool, collectively ensuring standardized text formatting regardless of input variations from different ASR systems.

\begin{figure}[!htbp]
\centering
\includegraphics[width=0.6\columnwidth, trim=0 0 0 0mm, clip]{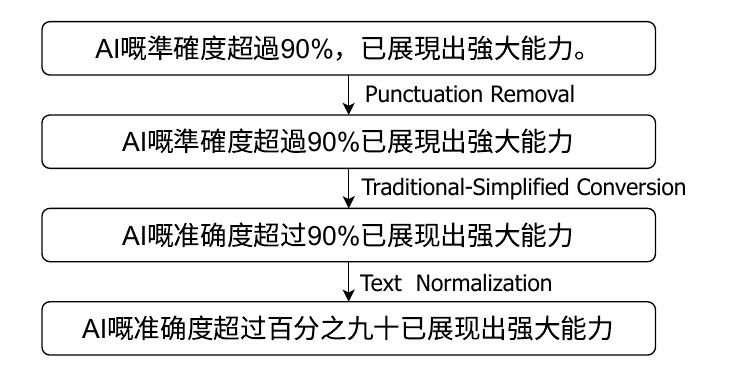} 
\caption{Text postprocessing example}
\label{text_processing}
\end{figure}


\subsection{LLM Corrector}

To enhance the accuracy of ASR transcriptions following multi-system voting, we implement an LLM-based correction module. The system employs \texttt{Qwen3-4B}\footnote{https://huggingface.co/Qwen/Qwen3-4B} as a dedicated Cantonese ASR correction expert, which processes the voted results alongside the original outputs from all three ASR systems as contextual references. This integrated approach generates refined transcriptions accompanied by confidence scores (0-100) and detailed correction analyses.

\begin{figure}[h]
\centering
\includegraphics[width=0.6\columnwidth]{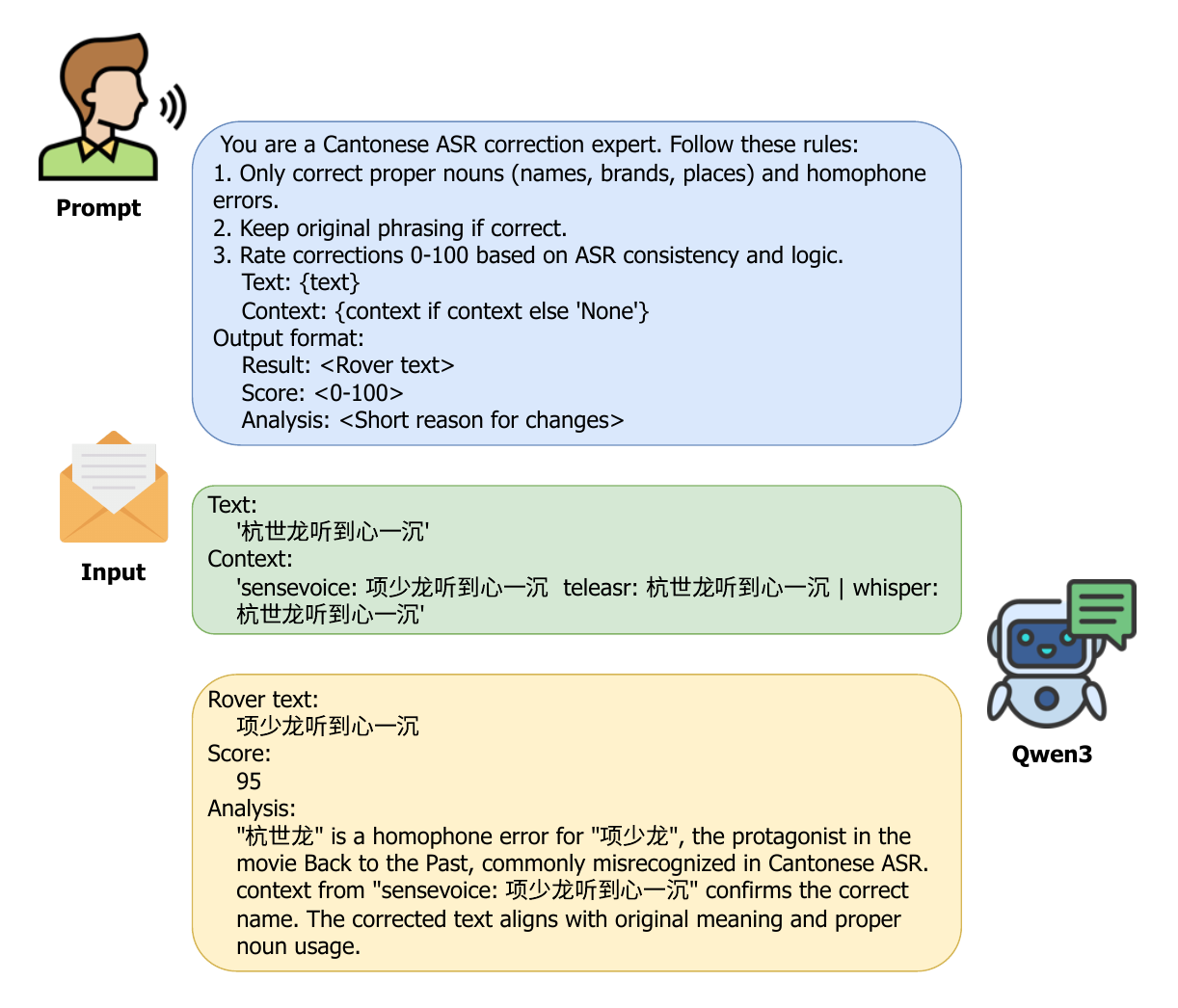} 
\caption{LLM corrector prompt}
\label{llm_corrector}
\end{figure}

\section{WSYue-TTS-eval Details}

In this section, we present the details of the Coverage Subset of WSYue-TTS-eval, including its construction process and representative examples.

\subsection{Coverage Subset Construction}

\begin{figure}[h]
\centering
\includegraphics[width=0.6\columnwidth]{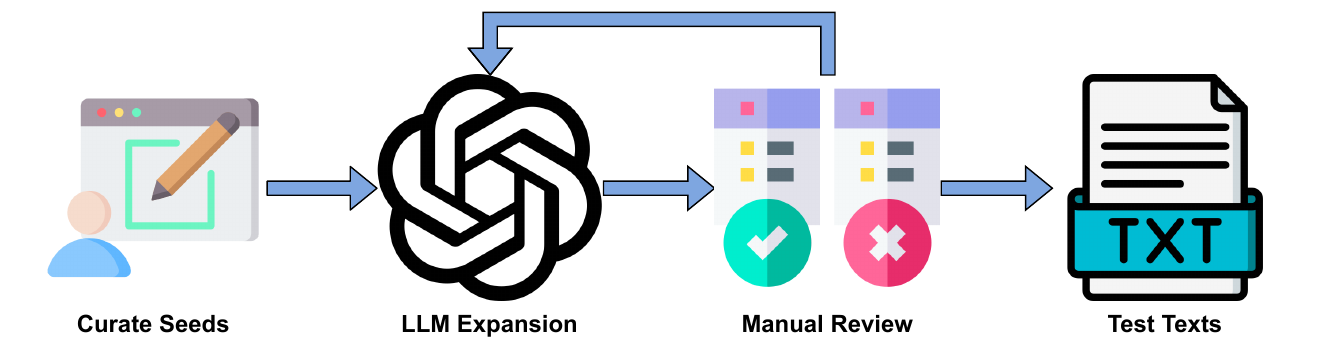}
\caption{Overview of the Coverage subset construction process.}
\label{fig:coverage_pipeline}
\end{figure}

\begin{figure}[h]
\centering
\includegraphics[width=0.6\columnwidth]{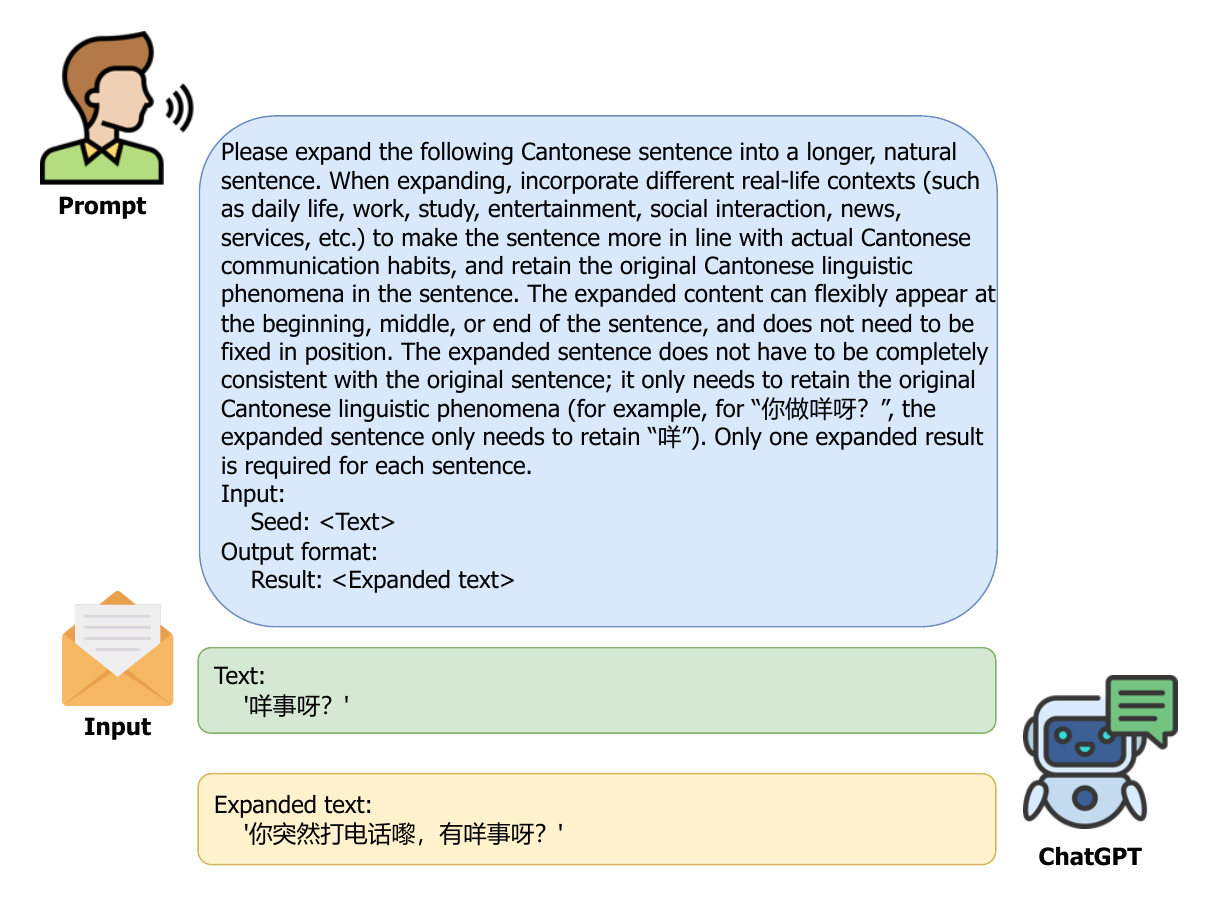}
\caption{The prompt provided to the large language model (LLM) for expanding seed texts into more natural and context-rich Cantonese sentences.}
\label{fig:llm_prompt}
\end{figure}


The \textit{Coverage} subset is constructed to systematically assess model performance across diverse domains and Cantonese linguistic phenomena. The overall construction pipeline is depicted in Figure~\ref{fig:coverage_pipeline} and comprises the following steps.


We first manually curated seed texts spanning multiple categories to ensure broad coverage. These categories include: (1) register and genre diversity (e.g., news, announcements, daily life, services, entertainment, culture, stories, and literature); (2) linguistic phenomena and structural complexity (e.g., numerals, named entities, polyphonic characters, code-switching, and strong emotional expressions); and (3) colloquial and conversational language (e.g., discourse particles, slang, and common expressions).

\begin{figure}[h]
\centering
\includegraphics[width=0.6\columnwidth]{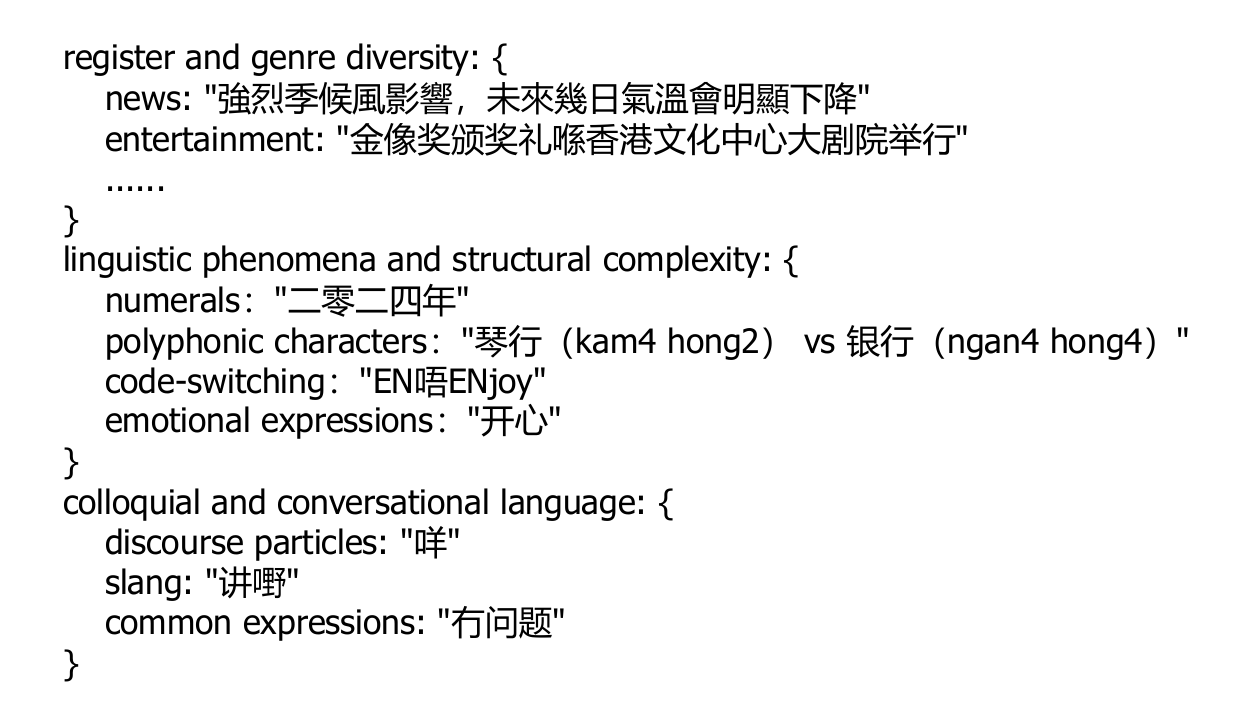} 
\caption{Examples of seed texts used in the Coverage subset.}
\label{fig:text_seed}
\end{figure}

For each category, we designed representative seed examples, as illustrated in Figure~\ref{fig:text_seed}. Some seed texts were already complete sentences and required no further expansion. For the remaining seeds, we utilized a large language model (LLM) to generate additional natural and fluent sentences suitable for TTS system input. The prompt provided to the LLM for sentence expansion is shown in Figure~\ref{fig:llm_prompt}.


All generated sentences were subsequently reviewed by human annotators to ensure correctness and adherence to Cantonese linguistic norms. 

\subsection{Coverage Subset Examples}
\begin{figure}[h]
    \centering
    \includegraphics[width=0.6\columnwidth]{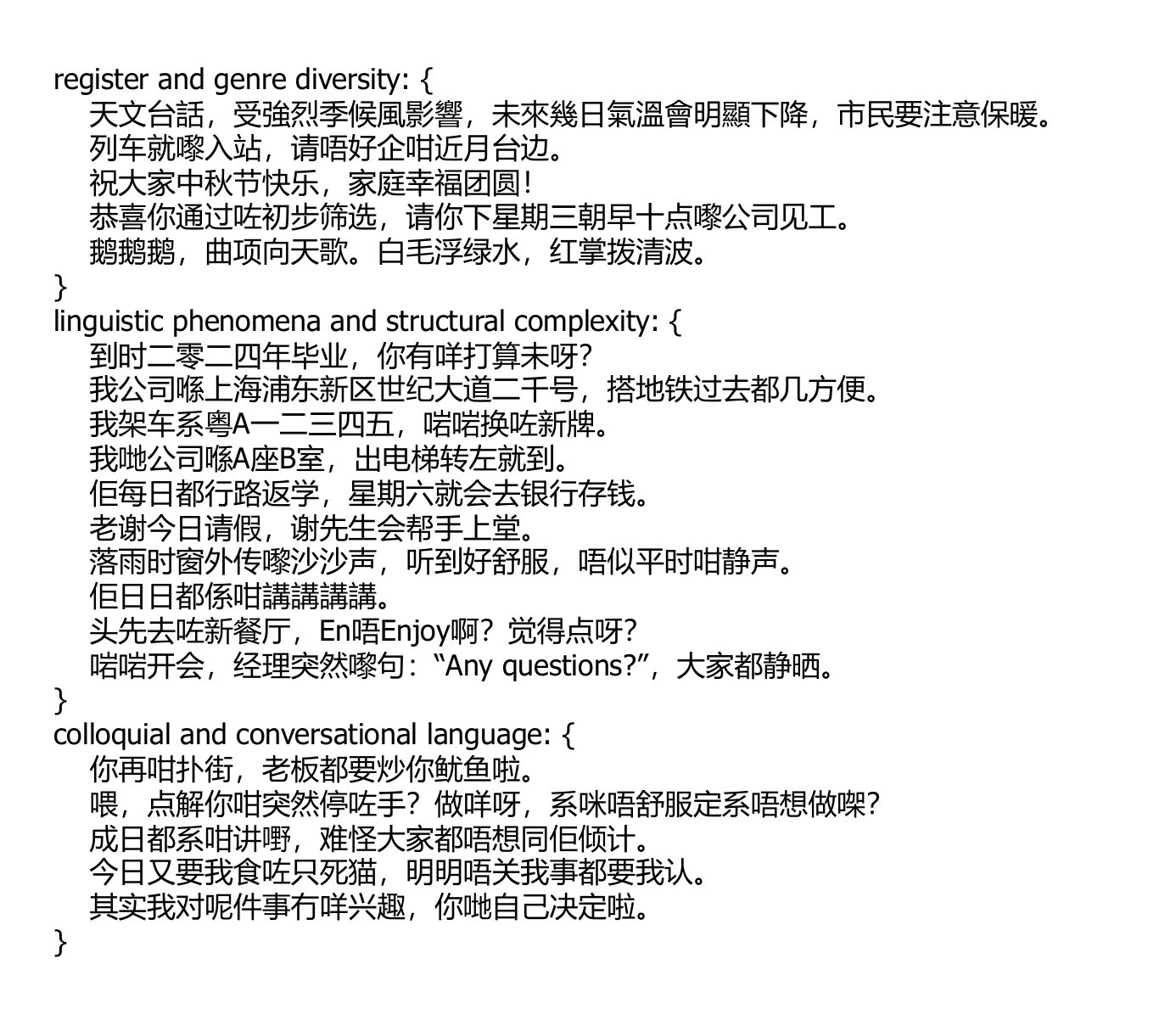}
    \caption{Representative examples from the Coverage Subset, covering various domains and Cantonese linguistic phenomena.}
    \label{fig:coverage_examples}
\end{figure}

To illustrate the diversity and representativeness of the Coverage Subset, we present several example sentences in Figure~\ref{fig:coverage_examples}. These examples cover a wide range of domains and linguistic phenomena, demonstrating the comprehensive nature of the dataset.

\end{document}